\documentclass[twocolumn,preprint]{aastex631}

\usepackage{CJK}
\usepackage{apjfonts}

\newcommand{\hal}{H$\alpha$}

\newcommand{\kms}{km s$^{-1}$}
\newcommand{\msun}{\ensuremath{M_\odot}}
\newcommand{\mbh}{\ensuremath{M_\mathrm{BH}}}
\newcommand{\per}{\ensuremath{^{-1}}}
\newcommand{\persq}{\ensuremath{^{-2}}}
\newcommand{\tcen}{\ensuremath{\tau_\mathrm{cen}}}
\newcommand{\rmax}{\ensuremath{r_\mathrm{max}}}
\newcommand{\tgz}{\ensuremath{\tau_\mathrm{gz}}}

\shorttitle{NGC 4395 Continuum Reverberation}
\shortauthors{Montano et al.}

\begin{document}
\begin{CJK}{UTF8}{gbsn}

\title{Optical Continuum Reverberation in the Dwarf Seyfert Nucleus of NGC 4395}

\author[0000-0001-5639-5484]{John W. Montano}
\affiliation{Department of Physics and Astronomy, 4129 Frederick Reines Hall, University of California, Irvine, CA, 92697-4575, USA}

\author[0000-0001-8416-7059]{Hengxiao Guo (郭恒潇)}
\affiliation{Department of Physics and Astronomy, 4129 Frederick Reines Hall, University of California, Irvine, CA, 92697-4575, USA}

\author[0000-0002-3026-0562]{Aaron J. Barth}
\affiliation{Department of Physics and Astronomy, 4129 Frederick Reines Hall, University of California, Irvine, CA, 92697-4575, USA}

\author[0000-0002-1912-0024]{Vivian U}
\affiliation{Department of Physics and Astronomy, 4129 Frederick Reines Hall, University of California, Irvine, CA, 92697-4575, USA}

\author[0000-0002-0164-8795]{Raymond Remigio}
\affiliation{Department of Physics and Astronomy, 4129 Frederick Reines Hall, University of California, Irvine, CA, 92697-4575, USA}

\author[0000-0002-9280-1184]{Diego H. Gonz\'{a}lez-Buitrago}
\affiliation{Universidad Nacional Aut\'onoma de M\'exico, Instituto de Astronom\'ia, AP 106,  Ensenada 22860, BC, M\'exico}

\author[0000-0002-6733-5556]{Juan V.\ Hern\'{a}ndez Santisteban}
\affiliation{SUPA School of Physics and Astronomy, North Haugh, St.~Andrews, KY16~9SS, Scotland, UK}

\begin{abstract}
The nearby dwarf spiral galaxy NGC 4395 contains a broad-lined active galactic nucleus (AGN) of exceptionally low luminosity powered by accretion onto a central black hole of very low mass ($\sim10^4-10^5$ \msun). In order to constrain the size of the optical continuum emission region through reverberation mapping, we carried out high-cadence photometric monitoring of NGC 4395 in the \emph{griz} filter bands on two consecutive nights in 2022 April using the four-channel MuSCAT3 camera on the Faulkes Telescope North at Haleakal\={a} Observatory. Correlated variability across the \emph{griz} bands is clearly detected, and the \emph{r}, \emph{i}, and \emph{z} band light curves show lags of $8.4^{+1.0}_{-1.1}$, $14.2^{+1.2}_{-1.4}$, and $20.4^{+2.0}_{-2.1}$ minutes with respect to the \emph{g} band when measured using the full-duration light curves. When lags are measured for each night separately, the Night 2 data exhibit lower cross-correlation amplitudes and shorter lags than the Night 1 light curves. Using the full-duration lags, we find that the lag-wavelength relationship is consistent with the $\tau\propto\lambda^{4/3}$ dependence found for more luminous AGN. Combining our results with continuum lags measured for other objects, the lag between $g$ and $z$ band scales with optical continuum luminosity as $\tgz \propto L^{0.56\pm0.05}$, similar to the scaling of broad-line region size with luminosity, reinforcing recent evidence that diffuse continuum emission from the broad-line region may contribute substantially to optical continuum variability and reverberation lags.

\end{abstract}


\section{Introduction} 
\label{sec:intro}

The active galactic nucleus (AGN) in the dwarf spiral galaxy NGC 4395 is the prototypical example of an AGN containing a very low-mass black hole (BH). First identified from the broad wings on its \hal\ emission line \citep{Filippenko1989}, it exhibits all of the characteristics of an accretion-powered Seyfert 1 nucleus, including a high-excitation photoionized narrow-line region \citep{Kraemer1999}, highly variable X-ray emission \citep{Lira1999, Iwasawa2000, Shih2003, Moran2005},  and a compact radio core \citep{Wrobel2001}, but its bolometric luminosity is just $\sim5\times10^{40}$ erg s\per\ \citep{Moran2005}, and the broad component of its \hal\ emission line has a dispersion of only $\sigma\approx600$ \kms\ \citep{Cho2021}. 
The mass of the BH in NGC 4395 is not precisely determined, but a variety of methods (including indirect estimates, reverberation mapping, and dynamical studies) point to a mass in the range $\sim10^4$ to a few $\times10^5$ \msun\ \citep{Kraemer1999, Shih2003, Filippenko2003, Peterson2005, denBrok2015, Woo2019, Cho2020}. While a growing number of AGN in this BH mass range have been detected in recent years \citep[for a review, see][]{Greene2020}, NGC 4395 remains uniquely amenable to detailed study due to its proximity at $d\approx4$ Mpc \citep{Thim2004}.

The small physical size of the AGN in NGC 4395 results in much shorter timescales for variability and reverberation at ultraviolet (UV) and optical wavelengths than are typical of luminous AGN. By fitting the TESS light curve of NGC 4395 with a damped random walk (DRW) model, \citet{Burke2020} found the DRW damping timescale to be $\tau_\mathrm{DRW} = 2.3_{-0.7}^{+1.8}$ days, in contrast to luminous AGN having $\tau_\mathrm{DRW}$ in the range of hundreds to thousands of days. Its broad emission-line reverberation lags are just $\sim$1 hour, both for the \ion{C}{4} $\lambda1549$ line measured relative to the 1350 \AA\ continuum \citep{Peterson2005} and for the \hal\ line measured relative to the $V$-band continuum \citep{Woo2019, Cho2020}.

Continuum reverberation mapping can provide a direct probe of the size of the UV/optical continuum emitting region in AGN, if the UV/optical continuum variations result from reprocessing of coronal emission by the accretion disk and/or gas in the broad-line region \cite[for a recent review see][]{Cackett2021RM}. While the continuum lags between UV and optical wavelengths in luminous Seyferts are typically a few days \citep[e.g.,][]{Edelson2019}, the continuum reverberation timescales in NGC 4395 are found to be much shorter. Using Hubble Space Telescope STIS UV observations from \citet{Peterson2005} combined with ground-based photometry, \citet{Desroches2006} measured a lag of $24_{-9}^{+7}$ minutes between the 1350 \AA\ continuum and the optical $V$ band.  \cite{McHardy2016} presented a preliminary report of an XMM-Newton and ground-based campaign on NGC 4395. Using data from the XMM-Newton Optical Monitor with the UVW1 filter ($\lambda_\mathrm{cen} = 2600$ \AA) and ground-based $g$-band photometry, they detected time delays of $473^{+47}_{-98}$ and $788^{+44}_{-54}$ s for the UVW1 and $g$ bands relative to the X-rays, which indicates that the lag of the $g$-band relative to the UVW1 band is $\sim$315 s.

Although time delays between UV and optical continuum bands in NGC 4395 have been measured, inter-band optical continuum lags have not been detected in prior work, largely due to the challenge of resolving very short timescales in observations that cycle between different filters on a single telescope \citep[e.g.,][]{Edri2012}. Extrapolating the 315 s time delay between 2600 \AA\ and the $g$ band \citep{McHardy2016} under the assumption of the $\tau\propto\lambda^{4/3}$ scaling relationship expected for reprocessing by a standard thin accretion disk \citep{Cackett2007}, the predicted lag between the $g$ and $z$ bands is just $\sim$11 minutes. The nightly $V$-band variability amplitude of the AGN is typically $\sim0.05-0.1$ mag \citep[e.g.,][]{Cho2020,Cho2021}, and an inter-band lag of several minutes is potentially detectable if photometry can be obtained with sufficiently high cadence and signal-to-noise ratio. Measurement of broad-band optical reverberation lags in NGC 4395 would provide new constraints on the size of its optical continuum emission region and a significant extension to the luminosity range of AGN having such constraints. In this \emph{Letter}, we report on the first unambiguous detection of continuum reverberation across the optical wavelength range in NGC 4395.

\section{Observations}
We observed NGC 4395 on 2022 April 26 and 27 (UT) using the MuSCAT3 camera \citep{Narita2020} on the 2 m Faulkes North Telescope (FTN) located at Haleakal\={a} Observatory. FTN is a part of the Las Cumbres Observatory (LCO) Global Telescope Network \citep{Brown2013}. MuSCAT3 is a four-channel imager that observes simultaneously in $g^\prime$, $r^\prime$, $i^\prime$, and $z_s$ filter bands (abbreviated as $griz$ hereinafter), with a $9\farcm1\times9\farcm1$ field of view and 0\farcs27 pixels. We used the fast readout mode (readout time 6 s) in order to obtain a rapid observing cadence, given the expectation that the inter-band lags might be as short as a few minutes. Exposure times were set to 100 s in the \emph{g}, \emph{i}, and \emph{z} bands. Since no suitable guide stars are available for use by the telescope autoguider in the field surrounding NGC 4395, we used the ``guide off'' mode, in which guiding is done using short exposures on one of the MuSCAT3 cameras. We chose the \emph{r} band for guiding, and used 25 s exposures in \emph{r}. On April 26 UT (Night 1), repeated observations in all four bands were carried out for a total duration of 6.7 hours, and on April 27 (Night 2) the observations spanned 6.2 hours. Conditions were mostly clear on both nights.

\begin{figure*}
 \centering
 \includegraphics[width=1\textwidth]{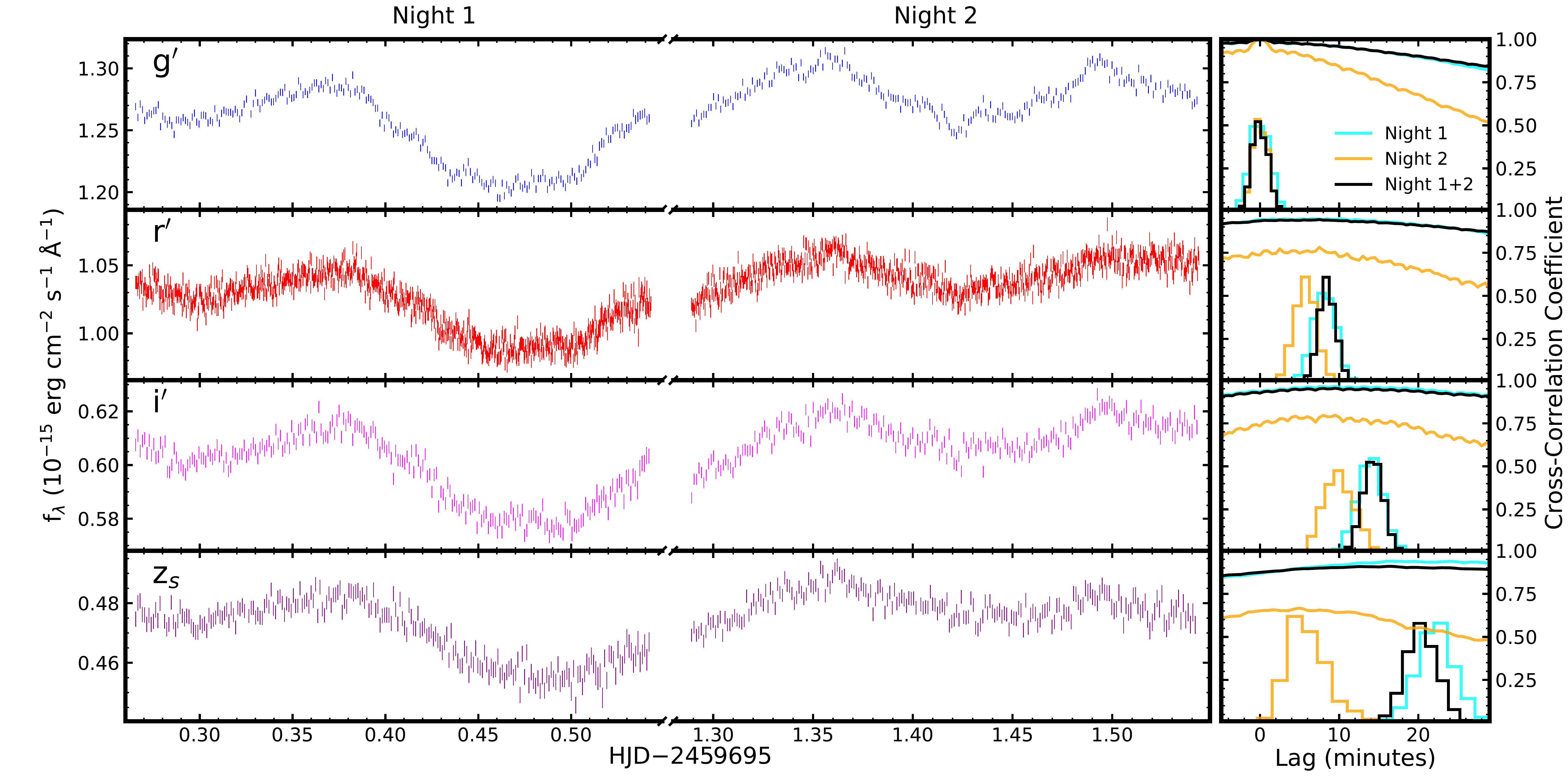}
    \caption{NGC 4395 light curves and cross-correlation lags. \emph{Left:} MuSCAT3 light curves in the \emph{griz} bands. \emph{Right:} Cross-correlation results for each band, measured relative to the \emph{g} band. Each panel shows cross-correlation functions (curves across the upper portion of the panel) and cross-correlation centroid distributions (histograms, renormalized to arbitrary units) for the data from Night 1 and from Night 2 separately, and for the combined Night 1 + Night 2 light curves. The \emph{g}-band panel displays the autocorrelation functions and autocorrelation centroid distributions of the \emph{g}-band light curves.} 
    \label{fig:lc}
\end{figure*}

\section{Data Reduction and Photometry}

The images were processed by the LCO BANZAI reduction pipeline, which includes bad pixel masking, bias subtraction, flat-field correction, and astrometric calibration using astrometry.net \citep{Lang2010}.

Photometry was carried out using an automated procedure based on Astropy \citep{Astropy2018} routines for aperture photometry and background sky measurement. In each image, the AGN and a set of comparison stars were identified by their coordinates. Aperture photometry was carried out using a photometric aperture radius of 3\arcsec, with the sky annulus spanning $10-20$\arcsec. We found that the best results were obtained using a single comparison star, SDSS J122550.91+333310.1, located close to the central region of NGC 4395. Counts were converted into flux density units using \emph{griz} photometric data obtained from the Sloan Digital Sky Survey (SDSS) for the comparison star. SDSS uses a $z^\prime$ filter while MuSCAT3 uses $z_s$, so the \emph{z}-band calibration is only approximate, but the overall normalization of the light curves does not affect the reverberation measurements.

The final light curves are displayed in Figure \ref{fig:lc} and the photometric data are listed in Table \ref{table:photometry}. Multiple features are present in the light curves, providing the required structure for cross-correlation lag measurement. The variations at longer wavelengths appear smoother and with lower amplitude than those in the $g$ band, as expected for a more extended reprocessing region at longer wavelengths, combined with the increasing contribution of light from the surrounding nuclear cluster \citep{Carson2015}.

\begin{deluxetable}{lCCC}
\tablecaption{NGC 4395 Photometry}
\tablehead{\colhead{Band} & \colhead{HJD$-2459695$} & \colhead{$f_\lambda$} & \colhead{$\sigma(f_\lambda)$}\\
\colhead{} &\colhead{} & \multicolumn{2}{c}{($10^{-15}$ erg cm\persq\ s\per\ \AA\per)}}
\startdata
\emph{g} & 0.2655 & 1.269 & 0.003 \\
\emph{g} & 0.2667 & 1.260 & 0.003 \\
\emph{g} & 0.2679 & 1.271 & 0.003 \\
\emph{g} & 0.2691 & 1.267 & 0.003 \\
\emph{g} & 0.2704 & 1.258 & 0.003 \\
\emph{g} & 0.2716 & 1.262 & 0.003 \\
\emph{g} & 0.2728 & 1.264 & 0.003\\
\emph{g} & 0.2740 & 1.263 & 0.003\\
\emph{g} & 0.2752 & 1.266 & 0.003\\
\emph{g} & 0.2765 & 1.266 & 0.003\\
\enddata
\label{table:photometry}
\tablecomments{ This table is published in its entirety in  machine-readable format. A portion is shown here for guidance regarding its form and content.}
\end{deluxetable}

\section{Lag Measurement}\label{sec:lag}

We measured the time lags of the \emph{riz} bands relative to the \emph{g} band using the interpolated cross-correlation function (ICCF) method as described by \citet{Gaskell1987} and \citet{White1994}, with the code PyCCF \citep{pyccf}. CCFs were measured over a lag search range spanning $-72$ to $72$ minutes, with a sampling interval of 0.5 minutes. The measurements were performed on the Night 1 and Night 2 data separately, and also on the combined Night 1 and 2 light curves. Error analysis was carried out using the standard flux randomization/random subset selection (FR/RSS) method \citep{Peterson1998}, with 5000 resampling iterations. In each iteration, the centroid of the CCF is determined from points above $0.8\rmax$, where \rmax\ is the peak height of the CCF. We take the lag for each band to be the median of the cross-correlation centroid distribution (CCCD) from the 5000 FR/RSS iterations.  The lags and \rmax\ values are listed in Table \ref{table:lags}.

Although the CCFs have very broad peaks, their centroids are well determined, with uncertainties of $\sim1-2$ minutes on the \tcen\ measurements. The Night 1 light curves yield very high \rmax\ values of 0.94--0.95, while the Night 2 data show substantially lower levels of inter-band correlation with \rmax\ values of 0.67--0.78. This can be attributed in part to a lower fractional variability amplitude during Night 2, but intrinsic changes in the continuum emission region structure may be partly responsible as well. The Night 2 lags are systematically shorter than those of Night 1, and the differences increase with wavelength. In particular, the \emph{z}-band lags exhibit a striking change, from 22.2 minutes on Night 1 to just 5.6 minutes on Night 2, and the difference between these is significant at the $4.6\sigma$ level considering the uncertainties on both measurements.  The lags measured from the combined Night 1 + Night 2 light curves are very similar to the Night 1 lags, with \rmax\ values nearly as high, indicating that the overall cross-correlation strength is dominated by the stronger variability in the Night 1 data. We take the combined Night 1 + 2 results as the final lag results for the following discussion.

\begin{deluxetable}{lCC|CC|CC}
\tablecaption{Continuum Reverberation Lags}
\tablehead{\colhead{Band} &  \multicolumn{2}{c}{Night 1} & 
\multicolumn{2}{c}{Night 2} & \multicolumn{2}{c}{Nights 1\&2} \\
            & \colhead{\tcen\ (min)} & \colhead{\rmax} 
            & \colhead{\tcen\ (min)} & \colhead{\rmax}     
            & \colhead{\tcen\ (min)} & \colhead{\rmax}}
\startdata  
\emph{g} (4770 \AA) & -0.03^{+1.21}_{-1.25} & 1.00 & 0.00^{+0.92}_{-0.92} & 1.00 & 0.04^{+0.98}_{-0.95} & 1.00 \\ 
\emph{r} (6215 \AA) & 8.14^{+1.44}_{-1.41} & 0.95 & 5.70^{+1.43}_{-1.30} & 0.78 & 8.44^{+0.99}_{-1.14} & 0.94 \\ 
\emph{i} (7545 \AA) & 13.89^{+1.67}_{-1.67} & 0.95 & 10.01^{+1.84}_{-2.02} & 0.80 & 14.20^{+1.20}_{-1.35} & 0.95\\ 
\emph{z} (8700 \AA) & 22.15^{+2.26}_{-2.42} & 0.94 & 5.63^{+2.09}_{-2.82} & 0.67 & 20.41^{+2.03}_{-2.05} & 0.91\\
\enddata
\label{table:lags}
\tablecomments{Lags are given in the observed frame in units of minutes. Central wavelengths are listed for each filter. All lags are measured relative to the \emph{g} band.}
\end{deluxetable}

\section{Discussion}

\begin{figure}
\hspace{-1cm}
 \centering
 \includegraphics[width=0.5\textwidth]{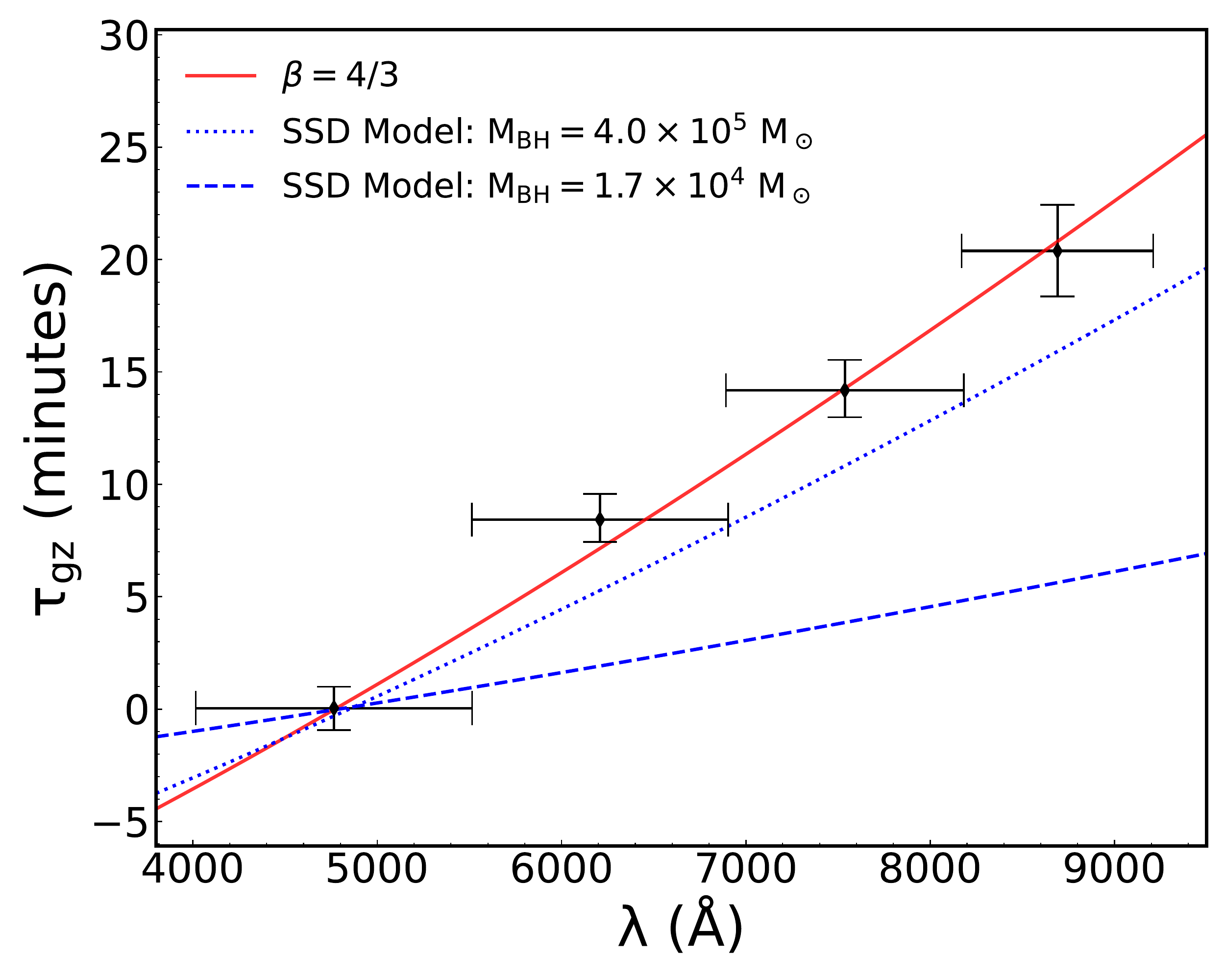}
    \caption{Lags as a function of wavelength in the rest frame of NGC 4395, from the combined Night 1 + Night 2 data. Horizontal error bars denote the width of each filter. The $\beta=4/3$ power-law fit to the \emph{riz} data points is shown as a red curve. Blue dashed and dotted curves represent SSD lag-wavelength models calculated for BH masses of $1.7\times10^4$ and $4\times10^5$ \msun, respectively.} 
    \label{fig:lagwave}
\end{figure}

Figure \ref{fig:lagwave} displays the inter-band lags corrected to the AGN rest frame (using $cz = 319$ \kms) as a function of rest-frame wavelength. To compare the data with disk reprocessing model predictions, we fit the \emph{riz} data points with the function $\tau(\lambda) = \tau_0 \left[ (\lambda/\lambda_0)^\beta - 1\right]$, where $\lambda_0$ is the central wavelength of the $g$-band filter shifted to the AGN rest frame ($\lambda_0 = 4765$ \AA), $\tau_0$ is a free parameter \citep[e.g.,][]{Edelson2019}, and we fix $\beta=4/3$ corresponding to a standard thin accretion disk. By construction, the model has $\tau=0$ at $\lambda=\lambda_0$, so we do not include the $g$-band data point in the fit. This simple model provides a good fit to the data, as shown in Figure \ref{fig:lagwave}, with $\chi^2_\nu = 0.91$ for two degrees of freedom.

We compare the observed lag-wavelength trend with model predictions for thermal reprocessing by a \citet{Shakura1973} disk (SSD) model following the method described by \citet{Fausnaugh2016} and \citet{Edelson2017}. In this model, the radius corresponding to peak emission at wavelength $\lambda$ is given by
\begin{equation}\label{eq:R}
   r(\lambda) =  \Bigg(X \frac{k\lambda}{hc} \Bigg)^{4/3} \Bigg[ \Bigg( \frac{GM_{\mathrm {BH}}}{8\pi\sigma} \Bigg)  \Bigg( \frac{L_{\mathrm{E}}}{\eta c^2}     \Bigg) (3+\kappa)\dot{m}   \Bigg]^{1/3},
\end{equation}
where $G$ is the gravitational constant, $M_{\rm BH}$ is the black hole mass, $\sigma$ is the Stefan-Boltzmann constant, $L_{\rm E}$ is the Eddington luminosity, $\dot{m}$ is the Eddington ratio, $\eta $ (set to 0.1) is the radiative efficiency, and $\kappa$ (set to 1) represents the relative contribution between X-rays and viscosity in disk heating. The quantity $X$ (set to 3.36) is a factor used in converting disk temperature to wavelength for a responsivity-weighted emitting radius \citep{Tie2018}. To calculate $\dot{m}$ we adopt the bolometric luminosity measurement of 5.3 $\times$ 10$^{40}$ erg s$\mathrm{^{-1}}$ from \citet{Moran2005}. Since the BH mass of NGC 4395 is highly uncertain, we calculate models for two values spanning the range of recent measurements: $\mbh=1.7\times10^4$ \msun\ from \hal\ reverberation mapping \citep{Cho2021} and $4\times10^{5}$ \msun\ from spatially resolved gas dynamics \citep{denBrok2015}. In Figure \ref{fig:lagwave} we show curves of the predicted lags relative to the $g$ band, $\tau(\lambda) = [r(\lambda)-r(\lambda_0)]/c$, for these two \mbh\ values.

Intensive UV/optical continuum reverberation mapping programs have found continuum emission region sizes that are typically $\sim2-3$ times larger than expected from disk reprocessing models \citep{Cackett2021RM}. While one possibility is that AGN accretion disks are larger than model predictions, recent work has focused on the contribution of diffuse continuum emission from the broad-line region (BLR) as an explanation for the longer-than-expected continuum lags and their wavelength dependence \citep[e.g.,][]{Cackett2018,Korista2019,Chelouche2019,Netzer2022}.
For NGC 4395 we find that the observed continuum emission region is 3.6 or 1.2 times larger than the disk reprocessing model predictions for BH masses of $1.7\times10^4$ or $4\times10^5$ \msun, respectively. These discrepancies would be $\sim50\%$ greater if $X=2.49$ was used; this value corresponds to a flux-weighted (rather than responsivity-weighted) emission radius and has been used in several other recent continuum reverberation studies \citep[e.g.,][]{Edelson2019}.  These model calculations for our chosen fiducial values of \mbh\ imply that NGC 4935 may have a ``disk size problem'' similar to that found in more luminous Seyferts, and if \mbh\ is at the lower end of the published range then NGC 4395 would be consistent with the trend identified by \citet{Li2021} in which lower-luminosity AGN exhibit larger disk size discrepancies. However, the standard disk reprocessing scenario could still be compatible with the data considering the BH mass uncertainty range of $\mbh=4_{-3}^{+8}\times10^5$ \msun\ ($3\sigma$ uncertainties) found by \citet{denBrok2015}.  Alternatively, these results could be interpreted as favoring a BH mass at the upper end of the likely range, if the basic disk reprocessing model was assumed to apply.

\begin{figure*}
 \centering
 \includegraphics[width=1\textwidth]{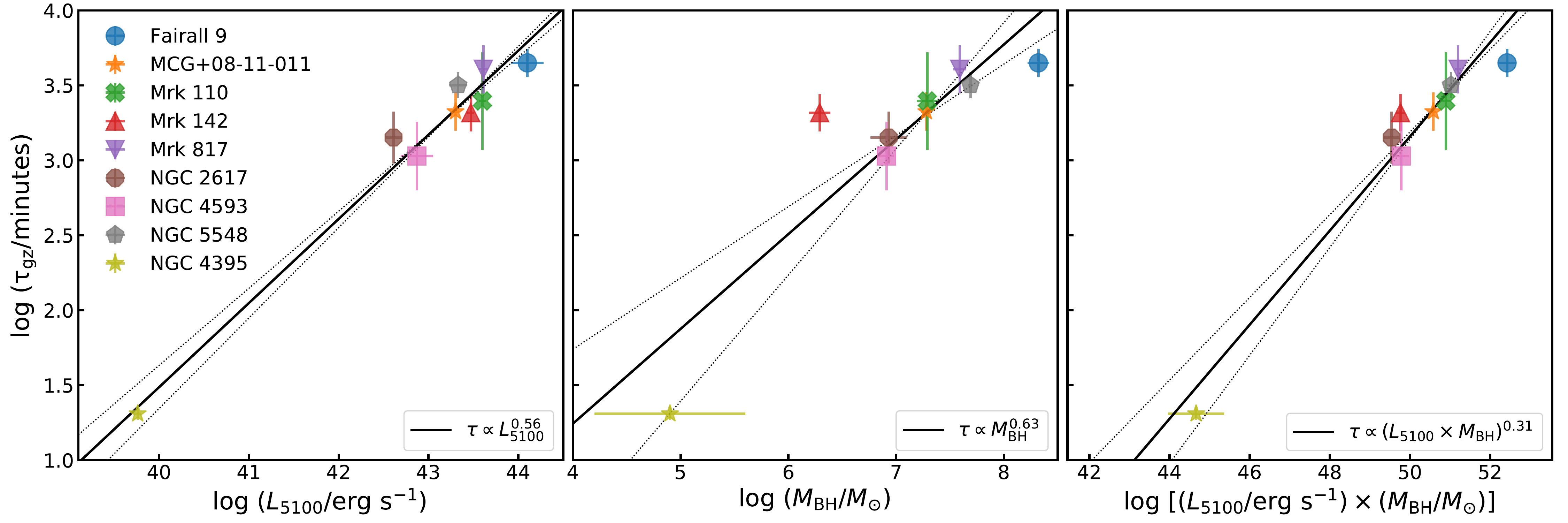}
    \caption{Rest-frame inter-band lag $\tau_{\rm gz}$ as a function of AGN continuum luminosity at 5100 \AA\ (\emph{left}), BH mass (\emph{middle}), and the product $L\times\mbh$ (\emph{right}). The $g-z$ lags of eight local AGN are from literature sources as described in the text. The AGN continuum luminosities and black hole masses for the comparison sample are from \citet{BentzKatz2015}. The quantity $L_{5100}$ refers to $\lambda L_\lambda$ at 5100 \AA. For NGC 4395, the horizontal error bar on \mbh\ denotes the range spanning $1.7\times10^4$ to $4\times10^5$ \msun\ corresponding to the mass measurements from \citet{Cho2021} and \citet{denBrok2015}. The black solid and dotted lines in each panel denote the power-law model fit and its 1$\sigma$ uncertainty range.}
    \label{fig:relation}
\end{figure*}

To compare the NGC 4395 continuum lags with these of more massive AGN, we selected eight local AGN (redshifts $0.017 < z < 0.047$) having high-quality \emph{g} through \emph{z}-band data from intensive disk reverberation mapping campaigns: Fairall 9 \citep{Hernandez2020}, MCG$+$08$-$11$-$011 \citep{Fausnaugh2018}, Mrk 110 \citep{Vincentelli2021}, Mrk 142 \citep{Cackett2020}, Mrk 817 \citep{Kara2021}, NGC 2617 \citep{Fausnaugh2018}, NGC 4593 \citep{Cackett2018}, and NGC 5548 \citep{Fausnaugh2016}. In these papers, the $g$ and $z$-band lags were measured relative to some other UV or optical band used as the driving light curve, but lags between $g$ and $z$ were not directly measured. In order to compare the measured \tgz\ of NGC 4395 with other AGN, we downloaded the light curve data from the original publications where available, and measured $\tau_\mathrm{gz}$ with the ICCF method as described previously.\footnote{For NGC 4593, the continuum light curves from \citet{Cackett2018} were measured from Hubble Space Telescope STIS spectroscopic data, and we used the published light curves at $\lambda=4745$ and 8800 \AA\ to approximate the \emph{g} and \emph{z} bands.} We performed a 2nd-order polynomial detrending for Fairall 9 before lag measurement to remove the long-term variability trend, following \citet{Hernandez2020}. For NGC 2617 and MCG$+$08$-$11$-$011 we used the best light-curve segments for lag measurement, following \citet{Fausnaugh2018}. For Mrk 110 and Mrk 817, the light curve data are not available in the published works, and we derived an estimate of \tgz\ by taking the difference between the published \emph{z}-band and \emph{g}-band lags, using error propagation to estimate the uncertainty on \tgz. 

The adopted observed-frame $\tau_{\mathrm{gz}}$ lags are 3.25$^{+0.79}_{-0.78}$ (Fairall 9), 1.5$^{+0.51}_{-0.51}$ (MCG$+$08$-$11$-$011), 1.79$^{+2.02}_{-1.07}$ (Mrk 110), 1.51$^{+0.50}_{-0.49}$ (Mrk 142), 2.90$^{+1.30}_{-1.34}$ (Mrk 817), 1.00$^{+0.49}_{-0.49}$ (NGC 2617), 0.75$^{+0.52}_{-0.46}$ (NGC 4593), and 2.24$^{+0.50}_{-0.50}$ (NGC 5548) days. AGN continuum luminosities at 5100 \AA\ and black hole masses for the comparison sample are available from the \citet{BentzKatz2015} catalog. For NGC 4395, we adopt the host-corrected AGN continuum luminosity of $\lambda L_\lambda = (5.75\pm0.4)\times10^{39}$ erg s$^{-1}$ at 5100 \AA\ from \citet{Cho2020}. As before, we consider a BH mass range from $1.7\times10^4$ to $4\times10^{5}$ \msun\ corresponding to recent measurements. We use the geometric mean of these values as the fiducial black hole mass with the uncertainty range spanning the range of these two values [$\log (M_{\mathrm{BH}}/\msun) = (4.9 \pm 0.7)$] in the following analysis.

Figure \ref{fig:relation} displays the relationship of \tgz\ with AGN continuum luminosity at 5100 \AA, and with \mbh. The 20.4 minute lag in NGC 4395 extends the dynamic range of the comparison sample downward by a factor of $\sim$50, and \tgz\ is clearly correlated with both luminosity and mass. The lag \tgz\ has a tighter relationship with $L$ than with \mbh, which could indicate that luminosity is a more fundamental driver of the continuum emission size, but larger uncertainties in the BH masses (particularly for NGC 4395) might also be responsible for the larger scatter in the \tgz-\mbh\ relation. 
We performed linear fits to the relations between $\log(\tgz/\mathrm{min})$ and $\log (L_{5100}/\mathrm{erg~s^{-1}})$ and between  $\log(\tgz/\mathrm{min})$ and  $\log (\mbh/\msun)$ with the code linmix \citep{Brandon2007}, which employs a hierarchical Bayesian model incorporating errors in both the luminosity and lag measurements. 

For the $\tgz-L$ relation we find $\log(\tgz/\mathrm{min}) = \left(0.56^{+0.05}_{-0.04}\right)\log (L_{5100}/\mathrm{erg~s^{-1}}) - \left(20.95^{+1.91}_{-1.97}\right)$, with a slope similar to the $r \propto L^{0.5}$ scaling of the broad-line region (BLR) radius-luminosity relationship \citep{Bentz2013}. This result is very similar to the recent finding by \citet{Netzer2022} that the $V$-band continuum lags of luminous Seyferts (measured relative to the UV) follow a $\tau\propto L^{0.5}_{5100}$ trend.  As argued by \citet{Netzer2022}, this observed lag-luminosity scaling suggests that diffuse continuum emission from the BLR \citep[e.g,][]{Korista2019} makes a substantial or even dominant contribution to the optical variability amplitude and continuum reverberation lags. The addition of NGC 4395 to the sample significantly extends the range of the $\tau-L$ relationship and demonstrates that the trend found by \citet{Netzer2022} continues to apply at the lowest AGN luminosities. This provides further corroborating evidence for the scenario in which BLR continuum emission contributes significantly to optical variability, as also demonstrated by the presence of a distinct excess lag at wavelengths below the Balmer jump in several objects \citep[e.g.,][]{Fausnaugh2018,Cackett2018,Edelson2019}.

For the $\tgz-\mbh$ relationship we obtain $\log(\tgz/\mathrm{min}) = \left(0.63^{+0.16}_{-0.21}\right) \log (\mbh/\msun) - \left(1.31^{+1.58}_{-1.15}\right)$; the slope of this relationship is not as tightly constrained due to the large uncertainty in the NGC 4395 BH mass.  

The disk model of Equation \ref{eq:R} predicts a scaling of $\tau \propto (L\times\mbh)^{1/3}$ at fixed $\lambda$. In the right panel of Figure \ref{fig:relation} we plot the relation between \tgz\ and the product $(L_{5100}\times\mbh)$ along with a power-law fit to the data with slope as a free parameter, again using $\log (M_{\mathrm{BH}}/\msun) = (4.9 \pm 0.7)$ to span the expected range in \mbh. The best fit gives $\tgz \propto (L\times\mbh)^{0.31\pm0.05}$, nearly identical to the disk reprocessing model prediction. However, this scaling can also be compatible with a scenario in which BLR diffuse continuum emission makes a large contribution to the optical variability, and where lag is primarily dependent on luminosity. Fitting a power law to the mass-luminosity relation of our sample yields $L_\mathrm{5100} \propto \mbh^{1.14}$. Combining the observed $\tgz \propto L_\mathrm{5100}^{0.56}$ relation with the mass-luminosity relation of the sample implies that the sample should follow $\tgz \propto (L_\mathrm{5100} \times \mbh)^{0.30}$, consistent with the observed  $\tgz \propto (L\times\mbh)^{0.31\pm0.05}$ scaling. In this way, a continuum lag-luminosity relation dominated by BLR reprocessing can produce a scaling of $\tau$ with $(L_{5100}\times\mbh)$ very similar to the disk reprocessing prediction. As a result, the $\tau$ vs.\ $(L\times\mbh)$ relation does not clearly discriminate between the disk and the BLR as the dominant location for continuum reprocessing in the optical spectrum, but observations of a larger sample spanning a broad range in Eddington ratio could potentially yield a useful test.

The difference in the measured lags between Night 1 and Night 2 indicates that the reprocessing geometry may undergo changes on very short timescales. The Night 2 data strongly depart from the expected lag-wavelength trend in that the \emph{z}-band lag is shorter than the \emph{i}-band lag, although the dominant reverberation behavior apparent in the combined Night 1+2 data follows the $\tau\propto\lambda^{4/3}$ model closely. Very few AGN have high-cadence continuum reverberation mapping data spanning timescales long enough to search for temporal changes in lags  \citep[one example is Mrk 110 which has shown evidence for a time-varying BLR contribution;][]{Vincentelli2022}. In light of this unusual behavior, it would be particularly useful to carry out monitoring of NGC 4395 over a longer time baseline, and over a broader range of wavelengths (ideally including X-ray and UV observations) to better explore temporal changes in lag behavior, and test whether variations in lag are correlated with any other observable properties. The dynamical timescale in the optically emitting region of the accretion disk in NGC 4395 will be of order days, compared with years for luminous Seyferts, making it possible to conduct studies of disk variability in NGC 4395 during a single observing season that would require years to decades to span the corresponding timescales in more typical AGN.

\section{Conclusions}
We have measured optical continuum reverberation lags in the dwarf AGN NGC 4395, finding lags of just $\sim8-20$ minutes for the \emph{riz} bands relative to the \emph{g} band. The trend of lag vs.\ wavelength is well matched by a $\tau \propto \lambda^{4/3}$ model, but the large uncertainty in BH mass precludes strong conclusions as to whether NGC 4395 has a ``disk size problem'' similar to that of high-luminosity AGN.  The addition of NGC 4395 extends the continuum lag-luminosity scaling found by \citet{Netzer2022} by two orders of magnitude in luminosity, and the derived slope of $\tau\propto L^{0.56}$ over this extended luminosity range further reinforces suggestions that diffuse BLR continuum emission contributes substantially to AGN optical continuum lags. Future continuum reverberation mapping of NGC 4395 over a broader wavelength range would be particularly valuable in order to search for evidence of diffuse continuum emission from the BLR, and to better understand short-timescale variations in reprocessing behavior.

\begin{acknowledgements}

This work makes use of observations from the Las Cumbres Observatory global telescope network. LCO telescope time was granted by NOIRLab through the Mid-Scale Innovations Program (MSIP). MSIP is funded by NSF. This paper is based on observations made with the MuSCAT3 instrument, developed by the Astrobiology Center and under financial supports by JSPS KAKENHI (JP18H05439) and JST PRESTO (JPMJPR1775), at Faulkes Telescope North on Maui, HI, operated by the Las Cumbres Observatory. We thank the staff of Las Cumbres Observatory and particularly Emily Manne-Nicholas for invaluable assistance in planning the MuSCAT3 observations. Research at UC Irvine was supported by NSF grant AST-1907290. DHGB acknowledges CONACYT support \#319800. JVHS acknowledges support from STFC grant ST/R000824/1.
We acknowledge Haleakal\={a} as a land of spiritual and cultural significance to the Native Hawaiian people. The use of this important site to further scientific knowledge is done so with appreciation and respect.
\end{acknowledgements}

\facility{FTN}
\software{AstroPy \citep{Astropy2018}, PyCCF \citep{pyccf}, Linmix \citep{Brandon2007}}

\bibliography{ms.bib}{}

\begin{thebibliography}{}
\expandafter\ifx\csname natexlab\endcsname\relax\def\natexlab#1{#1}\fi
\providecommand{\url}[1]{\href{#1}{#1}}
\providecommand{\dodoi}[1]{doi:~\href{http://doi.org/#1}{\nolinkurl{#1}}}
\providecommand{\doeprint}[1]{\href{http://ascl.net/#1}{\nolinkurl{http://ascl.net/#1}}}
\providecommand{\doarXiv}[1]{\href{https://arxiv.org/abs/#1}{\nolinkurl{https://arxiv.org/abs/#1}}}

\bibitem[{{Astropy Collaboration} {et~al.}(2018){Astropy Collaboration},
  {Price-Whelan}, {Sip{\H{o}}cz}, {G{\"u}nther}, {Lim}, {Crawford}, {Conseil},
  {Shupe}, {Craig}, {Dencheva}, {Ginsburg}, {VanderPlas}, {Bradley},
  {P{\'e}rez-Su{\'a}rez}, {de Val-Borro}, {Aldcroft}, {Cruz}, {Robitaille},
  {Tollerud}, {Ardelean}, {Babej}, {Bach}, {Bachetti}, {Bakanov}, {Bamford},
  {Barentsen}, {Barmby}, {Baumbach}, {Berry}, {Biscani}, {Boquien}, {Bostroem},
  {Bouma}, {Brammer}, {Bray}, {Breytenbach}, {Buddelmeijer}, {Burke},
  {Calderone}, {Cano Rodr{\'\i}guez}, {Cara}, {Cardoso}, {Cheedella}, {Copin},
  {Corrales}, {Crichton}, {D'Avella}, {Deil}, {Depagne}, {Dietrich}, {Donath},
  {Droettboom}, {Earl}, {Erben}, {Fabbro}, {Ferreira}, {Finethy}, {Fox},
  {Garrison}, {Gibbons}, {Goldstein}, {Gommers}, {Greco}, {Greenfield},
  {Groener}, {Grollier}, {Hagen}, {Hirst}, {Homeier}, {Horton}, {Hosseinzadeh},
  {Hu}, {Hunkeler}, {Ivezi{\'c}}, {Jain}, {Jenness}, {Kanarek}, {Kendrew},
  {Kern}, {Kerzendorf}, {Khvalko}, {King}, {Kirkby}, {Kulkarni}, {Kumar},
  {Lee}, {Lenz}, {Littlefair}, {Ma}, {Macleod}, {Mastropietro}, {McCully},
  {Montagnac}, {Morris}, {Mueller}, {Mumford}, {Muna}, {Murphy}, {Nelson},
  {Nguyen}, {Ninan}, {N{\"o}the}, {Ogaz}, {Oh}, {Parejko}, {Parley}, {Pascual},
  {Patil}, {Patil}, {Plunkett}, {Prochaska}, {Rastogi}, {Reddy Janga},
  {Sabater}, {Sakurikar}, {Seifert}, {Sherbert}, {Sherwood-Taylor}, {Shih},
  {Sick}, {Silbiger}, {Singanamalla}, {Singer}, {Sladen}, {Sooley},
  {Sornarajah}, {Streicher}, {Teuben}, {Thomas}, {Tremblay}, {Turner},
  {Terr{\'o}n}, {van Kerkwijk}, {de la Vega}, {Watkins}, {Weaver}, {Whitmore},
  {Woillez}, {Zabalza}, \& {Astropy Contributors}}]{Astropy2018}
{Astropy Collaboration}, {Price-Whelan}, A.~M., {Sip{\H{o}}cz}, B.~M., {et~al.}
  2018, \aj, 156, 123, \dodoi{10.3847/1538-3881/aabc4f}

\bibitem[{{Bentz} \& {Katz}(2015)}]{BentzKatz2015}
{Bentz}, M.~C., \& {Katz}, S. 2015, \pasp, 127, 67, \dodoi{10.1086/679601}

\bibitem[{{Bentz} {et~al.}(2013){Bentz}, {Denney}, {Grier}, {Barth},
  {Peterson}, {Vestergaard}, {Bennert}, {Canalizo}, {De Rosa}, {Filippenko},
  {Gates}, {Greene}, {Li}, {Malkan}, {Pogge}, {Stern}, {Treu}, \&
  {Woo}}]{Bentz2013}
{Bentz}, M.~C., {Denney}, K.~D., {Grier}, C.~J., {et~al.} 2013, \apj, 767, 149,
  \dodoi{10.1088/0004-637X/767/2/149}

\bibitem[{{Brown} {et~al.}(2013){Brown}, {Baliber}, {Bianco}, {Bowman},
  {Burleson}, {Conway}, {Crellin}, {Depagne}, {De Vera}, {Dilday}, {Dragomir},
  {Dubberley}, {Eastman}, {Elphick}, {Falarski}, {Foale}, {Ford}, {Fulton},
  {Garza}, {Gomez}, {Graham}, {Greene}, {Haldeman}, {Hawkins}, {Haworth},
  {Haynes}, {Hidas}, {Hjelstrom}, {Howell}, {Hygelund}, {Lister}, {Lobdill},
  {Martinez}, {Mullins}, {Norbury}, {Parrent}, {Paulson}, {Petry}, {Pickles},
  {Posner}, {Rosing}, {Ross}, {Sand}, {Saunders}, {Shobbrook}, {Shporer},
  {Street}, {Thomas}, {Tsapras}, {Tufts}, {Valenti}, {Vander Horst}, {Walker},
  {White}, \& {Willis}}]{Brown2013}
{Brown}, T.~M., {Baliber}, N., {Bianco}, F.~B., {et~al.} 2013, \pasp, 125,
  1031, \dodoi{10.1086/673168}

\bibitem[{{Burke} {et~al.}(2020){Burke}, {Shen}, {Chen}, {Scaringi},
  {Faucher-Giguere}, {Liu}, \& {Yang}}]{Burke2020}
{Burke}, C.~J., {Shen}, Y., {Chen}, Y.-C., {et~al.} 2020, \apj, 899, 136,
  \dodoi{10.3847/1538-4357/aba3ce}

\bibitem[{{Cackett} {et~al.}(2021){Cackett}, {Bentz}, \&
  {Kara}}]{Cackett2021RM}
{Cackett}, E.~M., {Bentz}, M.~C., \& {Kara}, E. 2021, iScience, 24, 102557,
  \dodoi{10.1016/j.isci.2021.102557}

\bibitem[{{Cackett} {et~al.}(2018){Cackett}, {Chiang}, {McHardy}, {Edelson},
  {Goad}, {Horne}, \& {Korista}}]{Cackett2018}
{Cackett}, E.~M., {Chiang}, C.-Y., {McHardy}, I., {et~al.} 2018, \apj, 857, 53,
  \dodoi{10.3847/1538-4357/aab4f7}

\bibitem[{{Cackett} {et~al.}(2007){Cackett}, {Horne}, \&
  {Winkler}}]{Cackett2007}
{Cackett}, E.~M., {Horne}, K., \& {Winkler}, H. 2007, \mnras, 380, 669,
  \dodoi{10.1111/j.1365-2966.2007.12098.x}

\bibitem[{{Cackett} {et~al.}(2020){Cackett}, {Gelbord}, {Li}, {Horne}, {Wang},
  {Barth}, {Bai}, {Bian}, {Carroll}, {Du}, {Edelson}, {Goad}, {Ho}, {Hu},
  {Khatu}, {Luo}, {Miller}, \& {Yuan}}]{Cackett2020}
{Cackett}, E.~M., {Gelbord}, J., {Li}, Y.-R., {et~al.} 2020, \apj, 896, 1,
  \dodoi{10.3847/1538-4357/ab91b5}

\bibitem[{{Carson} {et~al.}(2015){Carson}, {Barth}, {Seth}, {den Brok},
  {Cappellari}, {Greene}, {Ho}, \& {Neumayer}}]{Carson2015}
{Carson}, D.~J., {Barth}, A.~J., {Seth}, A.~C., {et~al.} 2015, \aj, 149, 170,
  \dodoi{10.1088/0004-6256/149/5/170}

\bibitem[{{Chelouche} {et~al.}(2019){Chelouche}, {Pozo Nu{\~n}ez}, \&
  {Kaspi}}]{Chelouche2019}
{Chelouche}, D., {Pozo Nu{\~n}ez}, F., \& {Kaspi}, S. 2019, Nature Astronomy,
  3, 251, \dodoi{10.1038/s41550-018-0659-x}

\bibitem[{{Cho} {et~al.}(2020){Cho}, {Woo}, {Hodges-Kluck}, {Son}, {Shin},
  {Gallo}, {Bae}, {Brink}, {Cho}, {Filippenko}, {Horst}, {Ili{\'c}}, {Joner},
  {Kang}, {Kang}, {Kaspi}, {Kim}, {Kova{\v{c}}evi{\'c}}, {Kumar}, {Le},
  {Nadzhip}, {Pozo Nu{\~n}ez}, {Metlov}, {Oknyansky}, {Park}, {Popovi{\'c}},
  {Rakshit}, {Schramm}, {Shatsky}, {Spencer}, {Sung}, {Sung}, {Tatarnikov}, \&
  {Vince}}]{Cho2020}
{Cho}, H., {Woo}, J.-H., {Hodges-Kluck}, E., {et~al.} 2020, \apj, 892, 93,
  \dodoi{10.3847/1538-4357/ab7a98}

\bibitem[{{Cho} {et~al.}(2021){Cho}, {Woo}, {Treu}, {Williams}, {Armen},
  {Barth}, {Bennert}, {Cho}, {Filippenko}, {Gallo}, {Geum},
  {Gonz{\'a}lez-Buitrago}, {G{\"u}ltekin}, {Hodges-Kluck}, {Horst}, {Hwang},
  {Kang}, {Kim}, {Kim}, {Leonard}, {Malkan}, {Remigio}, {Sand}, {Shin}, {Son},
  {Sung}, \& {U}}]{Cho2021}
{Cho}, H., {Woo}, J.-H., {Treu}, T., {et~al.} 2021, \apj, 921, 98,
  \dodoi{10.3847/1538-4357/ac1e92}

\bibitem[{{den Brok} {et~al.}(2015){den Brok}, {Seth}, {Barth}, {Carson},
  {Neumayer}, {Cappellari}, {Debattista}, {Ho}, {Hood}, \&
  {McDermid}}]{denBrok2015}
{den Brok}, M., {Seth}, A.~C., {Barth}, A.~J., {et~al.} 2015, \apj, 809, 101,
  \dodoi{10.1088/0004-637X/809/1/101}

\bibitem[{{Desroches} {et~al.}(2006){Desroches}, {Filippenko}, {Kaspi}, {Laor},
  {Maoz}, {Ganeshalingam}, {Li}, {Moran}, {Swift}, {Bentz}, {Ho}, {Nandra},
  {O'Neill}, \& {Peterson}}]{Desroches2006}
{Desroches}, L.-B., {Filippenko}, A.~V., {Kaspi}, S., {et~al.} 2006, \apj, 650,
  88, \dodoi{10.1086/507263}

\bibitem[{{Edelson} {et~al.}(2017){Edelson}, {Gelbord}, {Cackett}, {Connolly},
  {Done}, {Fausnaugh}, {Gardner}, {Gehrels}, {Goad}, {Horne}, {McHardy},
  {Peterson}, {Vaughan}, {Vestergaard}, {Breeveld}, {Barth}, {Bentz},
  {Bottorff}, {Brandt}, {Crawford}, {Dalla Bont{\`a}}, {Emmanoulopoulos},
  {Evans}, {Figuera Jaimes}, {Filippenko}, {Ferland}, {Grupe}, {Joner},
  {Kennea}, {Korista}, {Krimm}, {Kriss}, {Leonard}, {Mathur}, {Netzer},
  {Nousek}, {Page}, {Romero-Colmenero}, {Siegel}, {Starkey}, {Treu}, {Vogler},
  {Winkler}, \& {Zheng}}]{Edelson2017}
{Edelson}, R., {Gelbord}, J., {Cackett}, E., {et~al.} 2017, \apj, 840, 41,
  \dodoi{10.3847/1538-4357/aa6890}

\bibitem[{{Edelson} {et~al.}(2019){Edelson}, {Gelbord}, {Cackett}, {Peterson},
  {Horne}, {Barth}, {Starkey}, {Bentz}, {Brandt}, {Goad}, {Joner}, {Korista},
  {Netzer}, {Page}, {Uttley}, {Vaughan}, {Breeveld}, {Cenko}, {Done}, {Evans},
  {Fausnaugh}, {Ferland}, {Gonzalez-Buitrago}, {Gropp}, {Grupe}, {Kaastra},
  {Kennea}, {Kriss}, {Mathur}, {Mehdipour}, {Mudd}, {Nousek}, {Schmidt},
  {Vestergaard}, \& {Villforth}}]{Edelson2019}
---. 2019, \apj, 870, 123, \dodoi{10.3847/1538-4357/aaf3b4}

\bibitem[{{Edri} {et~al.}(2012){Edri}, {Rafter}, {Chelouche}, {Kaspi}, \&
  {Behar}}]{Edri2012}
{Edri}, H., {Rafter}, S.~E., {Chelouche}, D., {Kaspi}, S., \& {Behar}, E. 2012,
  \apj, 756, 73, \dodoi{10.1088/0004-637X/756/1/73}

\bibitem[{{Fausnaugh} {et~al.}(2016){Fausnaugh}, {Denney}, {Barth}, {Bentz},
  {Bottorff}, {Carini}, {Croxall}, {De Rosa}, {Goad}, {Horne}, {Joner},
  {Kaspi}, {Kim}, {Klimanov}, {Kochanek}, {Leonard}, {Netzer}, {Peterson},
  {Schn{\"u}lle}, {Sergeev}, {Vestergaard}, {Zheng}, {Zu}, {Anderson},
  {Ar{\'e}valo}, {Bazhaw}, {Borman}, {Boroson}, {Brandt}, {Breeveld}, {Brewer},
  {Cackett}, {Crenshaw}, {Dalla Bont{\`a}}, {De Lorenzo-C{\'a}ceres},
  {Dietrich}, {Edelson}, {Efimova}, {Ely}, {Evans}, {Filippenko}, {Flatland},
  {Gehrels}, {Geier}, {Gelbord}, {Gonzalez}, {Gorjian}, {Grier}, {Grupe},
  {Hall}, {Hicks}, {Horenstein}, {Hutchison}, {Im}, {Jensen}, {Jones},
  {Kaastra}, {Kelly}, {Kennea}, {Kim}, {Korista}, {Kriss}, {Lee}, {Lira},
  {MacInnis}, {Manne-Nicholas}, {Mathur}, {McHardy}, {Montouri}, {Musso},
  {Nazarov}, {Norris}, {Nousek}, {Okhmat}, {Pancoast}, {Papadakis}, {Parks},
  {Pei}, {Pogge}, {Pott}, {Rafter}, {Rix}, {Saylor}, {Schimoia}, {Siegel},
  {Spencer}, {Starkey}, {Sung}, {Teems}, {Treu}, {Turner}, {Uttley},
  {Villforth}, {Weiss}, {Woo}, {Yan}, \& {Young}}]{Fausnaugh2016}
{Fausnaugh}, M.~M., {Denney}, K.~D., {Barth}, A.~J., {et~al.} 2016, \apj, 821,
  56, \dodoi{10.3847/0004-637X/821/1/56}

\bibitem[{{Fausnaugh} {et~al.}(2018){Fausnaugh}, {Starkey}, {Horne},
  {Kochanek}, {Peterson}, {Bentz}, {Denney}, {Grier}, {Grupe}, {Pogge}, {De
  Rosa}, {Adams}, {Barth}, {Beatty}, {Bhattacharjee}, {Borman}, {Boroson},
  {Bottorff}, {Brown}, {Brown}, {Brotherton}, {Coker}, {Crawford}, {Croxall},
  {Eftekharzadeh}, {Eracleous}, {Joner}, {Henderson}, {Holoien}, {Hutchison},
  {Kaspi}, {Kim}, {King}, {Li}, {Lochhaas}, {Ma}, {MacInnis}, {Manne-Nicholas},
  {Mason}, {Montuori}, {Mosquera}, {Mudd}, {Musso}, {Nazarov}, {Nguyen},
  {Okhmat}, {Onken}, {Ou-Yang}, {Pancoast}, {Pei}, {Penny}, {Poleski},
  {Rafter}, {Romero-Colmenero}, {Runnoe}, {Sand}, {Schimoia}, {Sergeev},
  {Shappee}, {Simonian}, {Somers}, {Spencer}, {Stevens}, {Tayar}, {Treu},
  {Valenti}, {Van Saders}, {Villanueva}, {Villforth}, {Weiss}, {Winkler}, \&
  {Zhu}}]{Fausnaugh2018}
{Fausnaugh}, M.~M., {Starkey}, D.~A., {Horne}, K., {et~al.} 2018, \apj, 854,
  107, \dodoi{10.3847/1538-4357/aaaa2b}

\bibitem[{{Filippenko} \& {Ho}(2003)}]{Filippenko2003}
{Filippenko}, A.~V., \& {Ho}, L.~C. 2003, \apjl, 588, L13,
  \dodoi{10.1086/375361}

\bibitem[{{Filippenko} \& {Sargent}(1989)}]{Filippenko1989}
{Filippenko}, A.~V., \& {Sargent}, W. L.~W. 1989, \apjl, 342, L11,
  \dodoi{10.1086/185472}

\bibitem[{{Gaskell} \& {Peterson}(1987)}]{Gaskell1987}
{Gaskell}, C.~M., \& {Peterson}, B.~M. 1987, \apjs, 65, 1,
  \dodoi{10.1086/191216}

\bibitem[{{Greene} {et~al.}(2020){Greene}, {Strader}, \& {Ho}}]{Greene2020}
{Greene}, J.~E., {Strader}, J., \& {Ho}, L.~C. 2020, \araa, 58, 257,
  \dodoi{10.1146/annurev-astro-032620-021835}

\bibitem[{{Hern{\'a}ndez Santisteban} {et~al.}(2020){Hern{\'a}ndez
  Santisteban}, {Edelson}, {Horne}, {Gelbord}, {Barth}, {Cackett}, {Goad},
  {Netzer}, {Starkey}, {Uttley}, {Brandt}, {Korista}, {Lohfink}, {Onken},
  {Page}, {Siegel}, {Vestergaard}, {Bisogni}, {Breeveld}, {Cenko}, {Dalla
  Bont{\`a}}, {Evans}, {Ferland}, {Gonzalez-Buitrago}, {Grupe}, {Joner},
  {Kriss}, {LaPorte}, {Mathur}, {Marshall}, {Mehdipour}, {Mudd}, {Peterson},
  {Schmidt}, {Vaughan}, \& {Valenti}}]{Hernandez2020}
{Hern{\'a}ndez Santisteban}, J.~V., {Edelson}, R., {Horne}, K., {et~al.} 2020,
  \mnras, 498, 5399, \dodoi{10.1093/mnras/staa2365}

\bibitem[{{Iwasawa} {et~al.}(2000){Iwasawa}, {Fabian}, {Almaini}, {Lira},
  {Lawrence}, {Hayashida}, \& {Inoue}}]{Iwasawa2000}
{Iwasawa}, K., {Fabian}, A.~C., {Almaini}, O., {et~al.} 2000, \mnras, 318, 879,
  \dodoi{10.1046/j.1365-8711.2000.03810.x}

\bibitem[{{Kara} {et~al.}(2021){Kara}, {Mehdipour}, {Kriss}, {Cackett}, {Arav},
  {Barth}, {Byun}, {Brotherton}, {De Rosa}, {Gelbord}, {Hern{\'a}ndez
  Santisteban}, {Hu}, {Kaastra}, {Landt}, {Li}, {Miller}, {Montano},
  {Partington}, {Aceituno}, {Bai}, {Bao}, {Bentz}, {Brink}, {Chelouche},
  {Chen}, {Colmenero}, {Dalla Bont{\`a}}, {Dehghanian}, {Du}, {Edelson},
  {Ferland}, {Ferrarese}, {Fian}, {Filippenko}, {Fischer}, {Goad},
  {Gonz{\'a}lez Buitrago}, {Gorjian}, {Grier}, {Guo}, {Hall}, {Ho},
  {Homayouni}, {Horne}, {Ili{\'c}}, {Jiang}, {Joner}, {Kaspi}, {Kochanek},
  {Korista}, {Kynoch}, {Li}, {Liu}, {McHardy}, {McLane}, {Mitchell}, {Netzer},
  {Olson}, {Pogge}, {Popovi{\'c}}, {Proga}, {Storchi-Bergmann}, {Strasburger},
  {Treu}, {Vestergaard}, {Wang}, {Ward}, {Waters}, {Williams}, {Yang}, {Yao},
  {Zastrocky}, {Zhai}, \& {Zu}}]{Kara2021}
{Kara}, E., {Mehdipour}, M., {Kriss}, G.~A., {et~al.} 2021, \apj, 922, 151,
  \dodoi{10.3847/1538-4357/ac2159}

\bibitem[{{Kelly}(2007)}]{Brandon2007}
{Kelly}, B.~C. 2007, \apj, 665, 1489, \dodoi{10.1086/519947}

\bibitem[{{Korista} \& {Goad}(2019)}]{Korista2019}
{Korista}, K.~T., \& {Goad}, M.~R. 2019, \mnras, 489, 5284,
  \dodoi{10.1093/mnras/stz2330}

\bibitem[{{Kraemer} {et~al.}(1999){Kraemer}, {Ho}, {Crenshaw}, {Shields}, \&
  {Filippenko}}]{Kraemer1999}
{Kraemer}, S.~B., {Ho}, L.~C., {Crenshaw}, D.~M., {Shields}, J.~C., \&
  {Filippenko}, A.~V. 1999, \apj, 520, 564, \dodoi{10.1086/307486}

\bibitem[{{Lang} {et~al.}(2010){Lang}, {Hogg}, {Mierle}, {Blanton}, \&
  {Roweis}}]{Lang2010}
{Lang}, D., {Hogg}, D.~W., {Mierle}, K., {Blanton}, M., \& {Roweis}, S. 2010,
  \aj, 139, 1782, \dodoi{10.1088/0004-6256/139/5/1782}

\bibitem[{{Li} {et~al.}(2021){Li}, {Sun}, {Xu}, {Brandt}, {Trump}, {Yu},
  {Wang}, {Xue}, {Cai}, {Gu}, {Homayouni}, {Liu}, {Wang}, {Zhang}, \&
  {Li}}]{Li2021}
{Li}, T., {Sun}, M., {Xu}, X., {et~al.} 2021, \apjl, 912, L29,
  \dodoi{10.3847/2041-8213/abf9aa}

\bibitem[{{Lira} {et~al.}(1999){Lira}, {Lawrence}, {O'Brien}, {Johnson},
  {Terlevich}, \& {Bannister}}]{Lira1999}
{Lira}, P., {Lawrence}, A., {O'Brien}, P., {et~al.} 1999, \mnras, 305, 109,
  \dodoi{10.1046/j.1365-8711.1999.02388.x}

\bibitem[{{McHardy} {et~al.}(2016){McHardy}, {Connolly}, {Peterson}, {Bieryla},
  {Chand}, {Elvis}, {Emmanoulopoulos}, {Falco}, {Gandhi}, {Kaspi}, {Latham},
  {Lira}, {McCully}, {Netzer}, \& {Uemura}}]{McHardy2016}
{McHardy}, I.~M., {Connolly}, S.~D., {Peterson}, B.~M., {et~al.} 2016,
  Astronomische Nachrichten, 337, 500, \dodoi{10.1002/asna.201612337}

\bibitem[{{Moran} {et~al.}(2005){Moran}, {Eracleous}, {Leighly}, {Chartas},
  {Filippenko}, {Ho}, \& {Blanco}}]{Moran2005}
{Moran}, E.~C., {Eracleous}, M., {Leighly}, K.~M., {et~al.} 2005, \aj, 129,
  2108, \dodoi{10.1086/429522}

\bibitem[{{Narita} {et~al.}(2020){Narita}, {Fukui}, {Yamamuro}, {Harbeck},
  {Bowman}, {Elphick}, {Nation}, {Armstrong}, {Han}, {Abe}, {Ikoma}, {Isogai},
  {Kawauchi}, {Kurita}, {Kusakabe}, {de Leon}, {Livingston}, {Mori},
  {Nishiumi}, {Tamura}, {Watanabe}, {Volgenau}, {Heinrich-Josties}, {Foale},
  {Daily}, {McCully}, {Kirby}, {Smith}, {Haworth}, {Conway},
  {Storrie-Lombardi}, {Rosing}, {Chatelain}, {Bachelet}, {Johnson}, \&
  {Rabus}}]{Narita2020}
{Narita}, N., {Fukui}, A., {Yamamuro}, T., {et~al.} 2020, in Society of
  Photo-Optical Instrumentation Engineers (SPIE) Conference Series, Vol. 11447,
  Society of Photo-Optical Instrumentation Engineers (SPIE) Conference Series,
  114475K, \dodoi{10.1117/12.2559947}

\bibitem[{{Netzer}(2022)}]{Netzer2022}
{Netzer}, H. 2022, \mnras, 509, 2637, \dodoi{10.1093/mnras/stab3133}

\bibitem[{{Peterson} {et~al.}(1998){Peterson}, {Wanders}, {Horne}, {Collier},
  {Alexander}, {Kaspi}, \& {Maoz}}]{Peterson1998}
{Peterson}, B.~M., {Wanders}, I., {Horne}, K., {et~al.} 1998, \pasp, 110, 660,
  \dodoi{10.1086/316177}

\bibitem[{{Peterson} {et~al.}(2005){Peterson}, {Bentz}, {Desroches},
  {Filippenko}, {Ho}, {Kaspi}, {Laor}, {Maoz}, {Moran}, {Pogge}, \&
  {Quillen}}]{Peterson2005}
{Peterson}, B.~M., {Bentz}, M.~C., {Desroches}, L.-B., {et~al.} 2005, \apj,
  632, 799, \dodoi{10.1086/444494}

\bibitem[{{Shakura} \& {Sunyaev}(1973)}]{Shakura1973}
{Shakura}, N.~I., \& {Sunyaev}, R.~A. 1973, \aap, 24, 337

\bibitem[{{Shih} {et~al.}(2003){Shih}, {Iwasawa}, \& {Fabian}}]{Shih2003}
{Shih}, D.~C., {Iwasawa}, K., \& {Fabian}, A.~C. 2003, \mnras, 341, 973,
  \dodoi{10.1046/j.1365-8711.2003.06482.x}

\bibitem[{{Sun} {et~al.}(2018){Sun}, {Grier}, \& {Peterson}}]{pyccf}
{Sun}, M., {Grier}, C.~J., \& {Peterson}, B.~M. 2018, {PyCCF: Python Cross
  Correlation Function for reverberation mapping studies}, Astrophysics Source
  Code Library, record ascl:1805.032.
\newblock \doeprint{1805.032}

\bibitem[{{Thim} {et~al.}(2004){Thim}, {Hoessel}, {Saha}, {Claver}, {Dolphin},
  \& {Tammann}}]{Thim2004}
{Thim}, F., {Hoessel}, J.~G., {Saha}, A., {et~al.} 2004, \aj, 127, 2322,
  \dodoi{10.1086/382244}

\bibitem[{{Tie} \& {Kochanek}(2018)}]{Tie2018}
{Tie}, S.~S., \& {Kochanek}, C.~S. 2018, \mnras, 473, 80,
  \dodoi{10.1093/mnras/stx2348}

\bibitem[{{Vincentelli} {et~al.}(2022){Vincentelli}, {McHardy}, {Hern{\'a}ndez
  Santisteban}, {Cackett}, {Gelbord}, {Horne}, {Miller}, \&
  {Lobban}}]{Vincentelli2022}
{Vincentelli}, F.~M., {McHardy}, I., {Hern{\'a}ndez Santisteban}, J.~V.,
  {et~al.} 2022, \mnras, 512, L33, \dodoi{10.1093/mnrasl/slac009}

\bibitem[{{Vincentelli} {et~al.}(2021){Vincentelli}, {McHardy}, {Cackett},
  {Barth}, {Horne}, {Goad}, {Korista}, {Gelbord}, {Brandt}, {Edelson},
  {Miller}, {Pahari}, {Peterson}, {Schmidt}, {Baldi}, {Breedt}, {Hern{\'a}ndez
  Santisteban}, {Romero-Colmenero}, {Ward}, \& {Williams}}]{Vincentelli2021}
{Vincentelli}, F.~M., {McHardy}, I., {Cackett}, E.~M., {et~al.} 2021, \mnras,
  504, 4337, \dodoi{10.1093/mnras/stab1033}

\bibitem[{{White} \& {Peterson}(1994)}]{White1994}
{White}, R.~J., \& {Peterson}, B.~M. 1994, \pasp, 106, 879,
  \dodoi{10.1086/133456}

\bibitem[{{Woo} {et~al.}(2019){Woo}, {Cho}, {Gallo}, {Hodges-Kluck}, {Le},
  {Shin}, {Son}, \& {Horst}}]{Woo2019}
{Woo}, J.-H., {Cho}, H., {Gallo}, E., {et~al.} 2019, Nature Astronomy, 3, 755,
  \dodoi{10.1038/s41550-019-0790-3}

\bibitem[{{Wrobel} {et~al.}(2001){Wrobel}, {Fassnacht}, \& {Ho}}]{Wrobel2001}
{Wrobel}, J.~M., {Fassnacht}, C.~D., \& {Ho}, L.~C. 2001, \apjl, 553, L23,
  \dodoi{10.1086/320508}

\end{thebibliography}
\bibliographystyle{aasjournal}

\end{CJK}

\end{document}